\documentclass[11pt]{article}

\usepackage[margin=1in]{geometry}
\usepackage[T1]{fontenc}
\usepackage[utf8]{inputenc}
\usepackage{lmodern}
\usepackage{microtype}
\usepackage{setspace}
\usepackage{graphicx}
\usepackage{booktabs}
\usepackage{tabularx}
\usepackage{array}
\usepackage{ragged2e}
\usepackage{caption}
\usepackage{xcolor}
\usepackage{xurl}
\usepackage{hyperref}

\hypersetup{
  colorlinks=true,
  linkcolor=blue!45!black,
  citecolor=blue!45!black,
  urlcolor=blue!45!black,
  pdftitle={Agent-to-Agent Finance},
  pdfauthor={Hui Gong}
}

\newcolumntype{Y}{>{\RaggedRight\arraybackslash}X}
\title{\vspace{-1.5em}Agent-to-Agent Finance: Blockchain Payments and Trust Infrastructure for Autonomous AI Agents}
\author{Hui Gong\\{\small UCL Institute of Finance \& Technology}}
\date{August 2026}

\begin{document}
\maketitle

\begin{abstract}
Autonomous artificial intelligence (AI) agents are beginning to occupy a position between analytical tools and transacting counterparties. They can interpret goals, call external tools, negotiate with other agents, access data and computation, and in some settings initiate payments or blockchain transactions. This development creates a distinct problem for financial markets: if software agents can act economically under delegated authority, market participants need infrastructure for identity, authorisation, payment, verification, reputation and accountability. This chapter develops the concept of agent-to-agent finance as the layer of machine-mediated financial interaction in which autonomous agents discover counterparties, purchase services, express transaction intent, execute payments and generate auditable evidence. The argument is not that blockchain is a universal substrate for finance, but that programmable settlement, smart wallets, decentralised registries and verifiable computation can address specific coordination frictions created by autonomous agents. Drawing on recent work on blockchain agent-to-agent payments, ERC-8004 agent registries, provenance-based wallets, deterministic inference, decentralised finance (DeFi) intent mining, and official evidence on AI adoption in financial services, the chapter situates agent-to-agent finance as an emerging form of financial market infrastructure. It argues that the decisive design question is bounded autonomy: how to expand the set of economic actions that agents can safely perform without making markets more opaque, fragile or unaccountable.
\end{abstract}


\section{Introduction}

\subsection{Scope and motivation}

Financial markets have always depended on technical systems that translate intention into enforceable action. A quote becomes an order; an order becomes a trade; a trade becomes a cleared obligation; a cleared obligation becomes settlement; settlement becomes an auditable record. Artificial intelligence has entered these systems gradually, first as a statistical layer for prediction and classification, then as a workflow layer for surveillance, risk scoring, portfolio analytics and execution support. The current wave of agentic AI is different because it shifts the locus of action. Large language model (LLM)-based agents can plan, call tools, monitor state, respond to changing instructions and coordinate with other agents. When such systems are connected to wallets, application programming interfaces (APIs) and execution venues, they cease to be merely analytical instruments. They become delegated participants in economic exchange.

The motivation for studying agent-to-agent finance arises from this change in agency. Conventional AI in finance asks how models affect prediction, optimisation or decision support. Agentic finance asks how autonomous systems should be allowed to act. That distinction matters because financial markets are built around enforceability, responsibility and trust. A trading signal can be wrong without immediately transferring assets. A payment instruction, wallet signature or smart-contract transaction changes rights and obligations. Once an AI agent can initiate such actions, the market needs to know who authorised the action, what mandate constrained it, which counterparty was selected, whether the service was delivered, and how the action can be reconstructed after the fact.

Industry evidence suggests that the problem is no longer speculative. The Bank of England and the Financial Conduct Authority reported in 2024 that 75 percent of surveyed UK financial firms were already using AI, with a further 10 percent planning to use it within three years; 55 percent of all reported AI use cases had some degree of automated decision-making, even though only a small share were fully autonomous (Bank of England and FCA, 2024). That pattern is important. Financial institutions are not moving overnight to full automation. They are moving toward semi-autonomous systems embedded in operational processes, data procurement, compliance, risk analytics and customer service. Agentic AI will likely follow the same path: bounded autonomy in increasingly material workflows.

At the same time, the infrastructure around agents is developing rapidly. Google introduced Agent2Agent (A2A) as an open protocol for interoperable agents that can discover capabilities and coordinate tasks across enterprise systems (Google Developers Blog, 2025). Coinbase's x402 documentation presents programmable stablecoin payments over HTTP, including use cases in which AI agents pay for API access and digital services (Coinbase Developer Platform, n.d.). Blockchain research has begun to systematise agent-to-agent payments, agent identities, reputation registries, provenance-based wallets and verifiable AI outputs (Zhang et al., 2026; Xiong et al., 2026; Yu et al., 2026; Alves et al., 2026). These developments are not yet a coherent financial architecture, but they show the direction of travel: agents need ways to communicate, pay, authenticate, verify and account for economic actions.

The research gap is therefore not the absence of AI in finance, nor the absence of blockchain payment protocols. It is the lack of a financial-market theory of autonomous software actors. Existing AI-finance research often treats the model as a predictor. Existing DeFi research often treats automation as contract or bot execution. Existing payment research often treats the payor as a human, firm or account. Agent-to-agent finance sits between these literatures. It asks what happens when the operational payor is an adaptive software agent, the payment rail is programmable, and the counterparty may itself be another agent or machine service.

\subsection{Tokens, crypto-assets and the transaction boundary}

The central claim of this chapter is that agent-to-agent finance should be analysed as an emerging financial market infrastructure problem rather than as a narrow crypto use case or a generic AI application. The focus is not AI in general, nor artificial general intelligence (AGI), nor AI-generated content (AIGC) as a content-generation paradigm. The relevant object is the AI agent: a software system that can interpret a mandate, call tools, interact with counterparties and initiate economically meaningful actions. The key novelty is therefore not simply that AI agents may trade crypto-assets or tokenised financial instruments. It is that autonomous systems may become participants in service procurement, payment, data access, transaction routing, validation and reporting. This requires infrastructure for machine-native trust. Such infrastructure must bridge two different computational worlds: adaptive off-chain agents that reason under uncertainty, and deterministic on-chain or institutional systems that execute according to formal rules. The gap between these worlds is where many of the opportunities and risks arise.

This terminology also requires care. In AI, a token usually denotes a unit of text or computation consumed by a model. In blockchain and financial-market discussions, a token usually denotes a crypto-asset, tokenised claim, tokenised deposit, tokenised security or other ledger-based representation of value or rights. This chapter uses token only in the latter sense unless explicitly stated otherwise. The distinction matters because agentic finance may involve both forms at once: an agent can consume LLM tokens when reasoning, while using a wallet to pay stablecoins or interact with tokenised assets. Conflating the two obscures the difference between computational cost and financial settlement.

\subsection{Contribution and organising claim}

This chapter treats agent-to-agent finance as a field-level problem rather than a single protocol question. The conceptual model in Table 1 organises the field around four dimensions: actors, economic action, trust infrastructure and governance. The point of the model is to keep the analysis disciplined. Agentic finance is not every use of AI in finance; it concerns delegated software action that can create financial consequences and therefore requires machine-readable trust infrastructure.

\begin{table}[htbp]
\centering
\caption{A field-level conceptual model of agent-to-agent finance.}
\footnotesize
\renewcommand{\arraystretch}{1.16}
\begin{tabularx}{\textwidth}{YYYY}
\toprule
Dimension & Core question & Financial-market meaning & Examples in this chapter \\
\midrule
Actors & Who acts for whom? & The legal principal remains human or institutional, while operational action is delegated to software agents. & Principal, AI agent, model or tool provider, wallet operator and counterparty. \\
Economic action & What can the agent do? & A2A finance begins when an agent can create, route, pay for, settle, validate or evidence an economically relevant action. & Service discovery, data procurement, intent execution, payments, treasury operations and settlement workflows. \\
Trust infrastructure & How can the action occur safely? & Machine-native interaction requires identity, authorisation, payment, execution, verification and reputation. & Agent registries, smart wallets, policy controls, payment receipts, service-delivery metadata and audit logs. \\
Governance & Who remains accountable? & Delegated software action must be bounded by mandate, monitoring, escalation, recourse and supervisory evidence. & Know Your Agent, evidence envelopes, human escalation, liability allocation and regulatory observability. \\
\bottomrule
\end{tabularx}
\end{table}

The chapter makes four contributions. First, it defines agent-to-agent finance as machine-mediated financial interaction in which autonomous agents discover counterparties, express intent, initiate payments, execute transactions and generate evidence under delegated authority in settings such as data and model procurement, DeFi intent execution, autonomous treasury management, compliance checks, tokenised settlement workflows and machine-to-machine service markets. Second, it situates the concept within the evolution from Web3 automation toward Web4 agent economies, while avoiding the promotional language often associated with these labels. Third, it develops a trust-infrastructure framework that links agent identity, wallet authorisation, payment, execution, verification, reputation and governance. Fourth, it maps concrete financial applications and regulatory implications, showing why the topic matters for data markets, DeFi intent execution, autonomous treasury, compliance, tokenised settlement and supervisory observability.

The core proposition is that the economic value of agentic financial infrastructure comes not from maximising autonomy, but from expanding the set of actions that can be safely delegated. Bounded autonomy is therefore not a defensive afterthought. It is the design principle that makes agent-to-agent finance economically useful.

The practical concern running through the chapter is bounded autonomy. Financial institutions and market infrastructures do not need agents that can do anything. They need agents that can act within mandates, produce evidence, withstand manipulation and remain accountable. The design problem is therefore not how to maximise autonomy, but how to constrain it productively.

The remainder of the chapter proceeds as follows. Section 2 clarifies the concept and develops its theoretical foundations in delegation, transaction-cost economics and financial-market-infrastructure analysis. Section 3 frames agentic finance as an infrastructure upgrade from account-centric to agent-aware systems. Section 4 explains the blockchain trust rails that can support payments, wallet authorisation, identity and reputation. Section 5 examines verification, regulation, Know Your Agent and risk governance. Section 6 maps executable financial-market applications. Section 7 concludes by outlining a research agenda around identity, machine payments, accountability and systemic risk.

\section{Conceptual Foundations: From Financial Automation to Delegated Agentic Action}

\subsection{Evolution of financial automation}

The conceptual lineage of agent-to-agent finance begins before the current language of agentic AI. Algorithmic trading, robo-advice, smart contracts and decentralised finance (DeFi) bots all automate parts of financial activity. Algorithmic trading translates market signals into orders. Robo-advice automates portfolio allocation subject to suitability rules. Smart contracts execute state transitions when predefined conditions are met. decentralised finance (DeFi) bots monitor blockchain state and act on arbitrage, liquidation or rebalancing opportunities. These systems show that automation in finance is not new. What is new is the combination of autonomous planning, tool use, natural-language interfaces, cross-system interoperability and programmable payment.

Web3 contributed the idea that economic activity can be coordinated through wallets, tokens, smart contracts and shared ledgers. Its most mature applications were not the broad social visions attached to the term, but specific infrastructure primitives: custody, transfer, exchange, collateral, automated market making and programmable settlement. Web4, as used in recent agent-economy research, shifts the emphasis from user-owned assets to agent-operated services. In this framing, autonomous agents may hold wallets, pay for external API calls, execute on-chain trades and maintain verifiable identities (Jin et al., 2026). The term Web4 may or may not endure, but the underlying transition is analytically useful: the relevant actor is no longer only a human user interacting with a decentralised application, but a software agent acting across services on behalf of a principal.

The literature has moved through three stages. The first stage concerns blockchain automation: smart contracts, oracles, automated market makers and DeFi bots. These systems are rule-bound. They execute according to code or strategies, but they do not typically reason about open-ended tasks. The second stage concerns AI-assisted blockchain interaction: LLMs summarising transactions, detecting smart-contract vulnerabilities, explaining DeFi behaviour or generating trading signals. Here AI improves interpretation, but humans or scripts still control execution. The third stage concerns autonomous agent-blockchain interaction: agents observe state, infer intent, call tools, delegate tasks, sign transactions or pay counterparties. Recent surveys of autonomous agents on blockchains describe this as a progression from read-only analytics to delegated execution, autonomous signing and multi-agent workflows (Alqithami, 2026).

This is why the most immediate laboratory for agent-to-agent finance is likely to be DeFi rather than conventional regulated finance. In traditional finance, a software agent cannot simply appear as a counterparty: it must pass through institutional onboarding, account opening, know-your-customer (KYC) and anti-money-laundering (AML) screening, permissioning, outsourcing review and liability allocation. In DeFi, by contrast, wallets, smart contracts and public ledgers already permit software-controlled addresses to interact with financial protocols without a human-facing account layer. This does not mean DeFi is safer or normatively preferable; it means the technical surface for agentic experimentation already exists. A direct Web2-to-Web4 path is theoretically possible, but in practice it requires regulated identity, permission and accountability frameworks that are still being built.

Table 2 summarises this evolution from institutional automation and blockchain automation toward agent-mediated transactions. The table also shows why the newer problem is not simply automation itself, but the movement from rule-bound execution to delegated economic action.

\begin{table}[htbp]
\centering
\caption{Evolution from financial automation to agent-to-agent finance.}
\footnotesize
\renewcommand{\arraystretch}{1.16}
\begin{tabularx}{\textwidth}{YYYY}
\toprule
Stage & Typical systems & What is automated & Residual trust problem \\
\midrule
Institutional automation & Algorithmic trading, robo-advice, surveillance models & Signal generation, portfolio rules, execution support and monitoring inside regulated firms. & Model risk and oversight remain inside account-centric institutional wrappers. \\
Blockchain automation & Smart contracts, automated market makers, oracles and DeFi bots & Rule-bound execution and on-chain state transitions. & Code correctness, oracle integrity, key management and contract risk. \\
AI-assisted blockchain interaction & LLM explainers, smart-contract analysis and intent classifiers & Interpretation, summarisation and decision support. & Hallucination, data quality and human reliance. \\
Agent-blockchain interaction & Agent wallets, x402 payments and ERC-8004-style registries & Delegated discovery, payment, execution, validation and reconciliation. & Identity, authorisation, accountability and bounded autonomy. \\
\bottomrule
\end{tabularx}
\end{table}

\subsection{Definition and boundaries}

Agent-to-agent finance is narrower than the whole agent economy. It refers to transactions in which autonomous agents perform economically significant interactions. A simple chatbot that answers a question is not enough. A portfolio assistant that merely explains a risk report is not enough. An agent enters the domain of A2A finance when it can create or discharge obligations: paying for data, purchasing inference, routing an order, requesting settlement, validating a service, signing a transaction or producing an auditable compliance record. The concept therefore turns on economic action, not on conversational intelligence.

The definition also distinguishes A2A finance from conventional algorithmic trading. A trading algorithm is usually embedded in a specific institution, venue and strategy. Its risk controls are internal, and its counterparty relationships are mediated by brokers, exchanges or clearing members. A2A finance is more modular and service-oriented. One agent may purchase market data from another, request transaction analysis from a third, pay a solver for execution, and store an audit trail with a fourth service. The transaction chain becomes a network of agents and services rather than a single model connected to an order gateway.

This distinction matters for definitions. A financial agent should not be defined simply by autonomy or by the use of an LLM. A highly autonomous customer-service bot that cannot create obligations is outside the core of A2A finance. A simple script with wallet authority may be inside the perimeter if it can move assets or purchase services. The relevant boundary is the capacity to initiate or materially shape an economic transaction under delegated authority. This definition keeps the field analytically disciplined and avoids turning agentic finance into a catch-all for every AI system connected to a financial API.

A useful definition is therefore the following: agent-to-agent finance is the set of financial or quasi-financial arrangements in which autonomous software agents, operating under delegated mandates, discover counterparties, express transaction intent, exchange services, transfer value and generate verifiable records through machine-readable protocols. This definition is intentionally functional. It does not require every system to be decentralised, nor every payment to be on-chain. It asks what functions are necessary when autonomous systems become economic actors.

\subsection{Delegation, agency and bounded autonomy}

The theoretical core of agent-to-agent finance is delegation. In principal-agent theory, a principal delegates action to an agent when direct action is costly, incomplete or informationally difficult (Jensen and Meckling, 1976; Eisenhardt, 1989). Financial markets have long relied on such delegation: asset managers trade for clients, brokers route orders, custodians safeguard assets and payment institutions execute instructions. Agentic finance changes the operational form of the agent. The delegated actor is not necessarily a human professional or a tightly scripted system, but adaptive software that can interpret instructions, call tools and interact with counterparties.

This produces a layered delegation problem. A financial institution may delegate a task to an AI agent; the agent may rely on a foundation model, an external data provider, a wallet operator, a solver, an oracle or another agent; and each component may itself depend on software, cloud infrastructure or protocols controlled by third parties. The traditional agency problem of monitoring an agent's behaviour is therefore extended into a chain of operational agency. The key governance question is no longer only whether a person or account was authorised, but whether every economically material step remained inside the principal's mandate.

Transaction-cost economics provides a second foundation. Firms and markets exist partly because search, bargaining, monitoring and enforcement are costly (Coase, 1937; Williamson, 1979). Agents can lower the marginal cost of discovering digital services, contracting through standard interfaces, paying for small tasks and reconciling evidence. At the same time, they introduce new agency costs, verification costs and governance costs. The economic problem is therefore a trade-off between automation gains and trust costs. A2A finance is valuable only when infrastructure reduces the combined cost of coordination, verification and accountability.

A third foundation comes from financial market infrastructure. Market infrastructure is not merely technical plumbing; it provides finality, standardisation, risk controls and records that allow financial obligations to be trusted at scale (Committee on Payments and Market Infrastructures and IOSCO, 2012; Merton and Bodie, 1995). Agent-to-agent finance belongs in this tradition because autonomous agents create a new object of infrastructure: delegated software action. The chapter therefore uses bounded autonomy as an organising proposition: the objective is maximum safely delegable economic action, not maximum machine discretion.

\subsection{From Web3 infrastructure to agent economies}

The transition from Web3 to Web4 should also be understood as a change in the architecture of demand. In Web3, a human user normally decides to swap, lend, mint or bridge, and the protocol executes. In an agent economy, the demand for a service may be generated by software. A research agent may decide it needs an additional dataset; a treasury agent may decide it needs liquidity; a compliance agent may decide it needs a sanctions check; a solver may decide it needs a price feed. The economic demand is still ultimately attributable to a human or institution, but the immediate initiator is an agent. That creates a new class of machine-originated financial instructions.

\section{Agentic Finance as Financial Market Infrastructure}

\subsection{From account-centric to agent-aware infrastructure}

The financial-market significance of agent-to-agent finance becomes clearer when it is framed as infrastructure. Financial market infrastructure (FMI) traditionally refers to the systems that enable trading, clearing, settlement, custody, payments and record-keeping. Exchanges organise price discovery. Central counterparties manage counterparty risk. Central securities depositories record ownership and settlement. Payment systems move money. Custodians safeguard assets. Messaging networks transmit instructions. Regulators and supervisors rely on reporting systems to reconstruct activity. These infrastructures are not background plumbing; they define what market participants can do and how trust is maintained. The international principles for financial market infrastructures emphasise legal basis, governance, credit and liquidity risk, settlement finality, operational risk and transparency (Committee on Payments and Market Infrastructures and IOSCO, 2012).

Traditional financial infrastructure was designed around identifiable legal persons, regulated intermediaries and relatively stable institutional boundaries. Even when trading became electronic, the actors remained firms, accounts, brokers, clearing members and custodians. Algorithmic systems acted inside these legal wrappers. Agentic finance weakens that neat separation. An agent may be deployed by a firm but rely on a third-party model; it may use a wallet operated by another provider; it may purchase data from a marketplace; it may call a solver outside the institution; it may execute on a public blockchain or a tokenised deposit platform. The legal principal remains human or institutional, but the operational actor becomes distributed across software components.

This is why financial-market infrastructure must evolve from account-centric automation to agent-aware automation. In an account-centric system, the question is whether an account is authorised to act. In an agent-aware system, the question becomes more granular: which agent acted, under which mandate, using which tool, with which wallet permission, on whose behalf, and with what evidence. The same account may host several agents with different scopes. The same agent may act for different principals. The same principal may delegate tasks to multiple agents. Infrastructure must therefore support attribution at the level of delegated software action.

The reason this infrastructure is needed is not technological novelty for its own sake. Traditional financial systems were designed around accounts, firms and human-authorised workflows. Agentic finance inserts a new operational actor between the principal and the transaction: a software agent that can select services, spend budget, request execution and generate records. Unless infrastructure evolves, firms face a control gap: the legal person remains accountable, but the operational decision chain is distributed across models, tools, wallets and service providers.

Table 3 clarifies the boundary between the legal principal and the operational actor. This distinction matters because the chapter does not argue that an AI agent becomes a legal person. It argues that financial infrastructure must recognise operational agency without dissolving legal responsibility.

\begin{table}[htbp]
\centering
\caption{Legal principal and operational actor in agentic finance.}
\footnotesize
\renewcommand{\arraystretch}{1.16}
\begin{tabularx}{\textwidth}{YYYY}
\toprule
Layer & Traditional account-centric infrastructure & Agent-aware infrastructure & Design implication \\
\midrule
Legal subject & Customer, firm, broker, custodian or account holder. & The same legal person remains responsible for the delegated action. & A2A finance does not require treating the agent as an independent legal person. \\
Operational actor & Human employee, rule-based system or internal trading algorithm. & Delegated software agent using models, tools, wallets and external services. & Infrastructure must attribute action below the level of the account. \\
Authority object & Account permission, user credential, trading limit or API key. & Mandate, agent identity, wallet scope, policy version and evidence record. & Capability must be separated from permission. \\
Audit question & Did the authorised account or system perform the transaction? & Which agent acted, under which mandate, using which tool and with what evidence? & Logs must reconstruct the delegated decision chain. \\
\bottomrule
\end{tabularx}
\end{table}

\subsection{Why existing infrastructure is being forced to upgrade}

The current upgrade of financial infrastructure already points in this direction. Open banking and open finance make data access more API-driven. Tokenisation projects attempt to represent deposits, securities or real-world assets on programmable ledgers. Wholesale payment experiments such as Bank for International Settlements (BIS) Project Agora explore shared programmable platforms for tokenised central bank reserves and commercial bank deposits, with atomic settlement and embedded workflow logic (Bank for International Settlements, 2026). Cloud infrastructure and foundation models make financial services more modular and dependent on third parties. Agentic AI sits on top of these trends. It is not an isolated technology; it is a new control layer for a more programmable financial system.

The Bank of England and FCA survey is instructive. It reports that AI adoption is widespread, that a third of AI use cases are third-party implementations, and that the top three model providers account for 44 percent of all named model providers (Bank of England and FCA, 2024). These figures matter for agent-to-agent finance because they show that the next generation of financial automation will not be built wholly inside single firms. It will depend on external models, cloud providers, data vendors, payments APIs and specialised services. Agentic workflows will intensify the need to manage third-party dependencies because agents will not merely consume third-party services; they may select, pay and coordinate them.

\subsection{Hybrid market infrastructure}

This vision should not be confused with full disintermediation. In regulated finance, banks, brokers, custodians, asset managers and market infrastructures will remain central. The question is how their systems incorporate autonomous agents. A bank may not allow an external agent to sign transactions directly, but it may use agents to prepare instructions, gather evidence or monitor exceptions. A DeFi protocol may allow more direct agent participation, but still need identity, reputation and validation. A tokenised deposit platform may permit conditional payments, but require compliance checks and legal finality. Agent-to-agent finance therefore spans a continuum from internal institutional automation to open agent economies.

The market-infrastructure framing also clarifies the economic value. Agents can reduce search costs by discovering services dynamically; reduce transaction costs by automating payment and reconciliation; reduce information costs by purchasing and processing data; reduce coordination costs by interacting across protocols; and reduce monitoring costs by producing structured logs. Yet the same capabilities can increase systemic opacity if many institutions rely on similar agents, model providers or data sources. The Financial Stability Board (FSB) has warned that AI adoption may amplify vulnerabilities through third-party dependencies, market correlations, cyber risk, model risk and governance failures (FSB, 2024). Agent-to-agent finance brings those vulnerabilities closer to payment and settlement.

\section{Blockchain Trust Rails for Autonomous Agents}

\subsection{Adaptive decision formation and deterministic settlement}

The role of blockchain in agent-to-agent finance should be framed narrowly and precisely. Blockchain does not make agents trustworthy. It does not solve model risk, prompt injection, legal responsibility or poor governance. Its relevance lies in a subset of functions that autonomous agents require: programmable payments, persistent identifiers, auditable state, smart-account policy, composable execution and, in some designs, cryptoeconomic validation. These functions can reduce coordination frictions when agents transact with unknown or semi-trusted counterparties.

A useful way to understand the problem is to compare adaptive AI agents with deterministic blockchain machinery. A Bitcoin miner follows a protocol. It selects transactions, constructs blocks and performs proof-of-work according to rules that are fixed relative to the protocol state. It does not interpret a user's broad economic mandate or decide whether a service is reputable. By contrast, an AI agent may update its plan, interpret natural language, query external tools, respond to incomplete information and choose among counterparties. This adaptiveness creates value, but it also makes the agent less predictable than a miner or smart contract. The agent's reasoning is probabilistic and context-dependent; blockchain execution is deterministic and state-dependent. Agent-to-agent finance must connect these two forms of computation without pretending they are the same.

Table 4 develops this comparison. The point is not to make the Bitcoin miner the theoretical anchor of the chapter, but to show why adaptive agents need a bridge to deterministic settlement and policy enforcement.

\begin{table}[htbp]
\centering
\caption{Deterministic blockchain machinery versus adaptive AI agents.}
\footnotesize
\renewcommand{\arraystretch}{1.16}
\begin{tabularx}{\textwidth}{YYYY}
\toprule
Dimension & Bitcoin miner or smart contract & Autonomous financial agent & Design implication \\
\midrule
Logic & Executes protocol-defined rules. & Interprets goals and revises plans under uncertainty. & Formal execution must be paired with policy constraints. \\
Inputs & Transactions and blockchain state. & Prompts, APIs, tools, market data, memories and counterparty claims. & Input provenance and tool permissions become central. \\
Failure mode & Protocol bug, key compromise or economic attack. & Prompt injection, hallucination, mandate drift, tool misuse or manipulated confidence. & Security must cover semantic as well as cryptographic attacks. \\
Accountability & Attribution to address, contract or miner behaviour. & Attribution across principal, operator, model, wallet, policy layer and counterparty. & Logs must bind technical action to delegated authority. \\
\bottomrule
\end{tabularx}
\end{table}

This distinction explains why smart contracts alone are insufficient. A smart contract can enforce an escrow condition, spending limit or settlement rule. It cannot by itself determine whether an LLM agent was manipulated by a malicious web page, whether a data service delivered economically meaningful content, whether an agent's action remained within an institutional mandate, or whether a reputation score reflects genuine performance. The trust infrastructure must therefore include off-chain controls, on-chain commitments, wallet policies, logs and governance procedures.

\subsection{A lifecycle trust stack}

The eight-layer trust stack in Figure 1 is best understood as a lifecycle model. Discovery, identity and authorisation are pre-transaction trust functions: they answer who or what may interact, and under whose authority. Payment and execution are transaction trust functions: they translate economic intent into a state change. Verification, reputation and governance are post-transaction trust functions: they determine whether the outcome is evidenced, learned from, challenged and remedied.

The trust stack proposed here has eight layers: discovery, identity, authorisation, payment, execution, verification, reputation and governance. Discovery allows an agent to find services or counterparties. Identity links an agent to an operator, wallet, service endpoint or credential. Authorisation specifies what the agent may do. Payment transfers value. Execution calls tools, APIs or smart contracts. Verification checks whether outcomes correspond to claims. Reputation accumulates structured evidence about past behaviour. Governance resolves failures and allocates responsibility. The stack is not a technology list; it is a set of functions that any serious A2A financial system must implement.

\begin{figure}[htbp]
\centering
\includegraphics[width=0.96\textwidth]{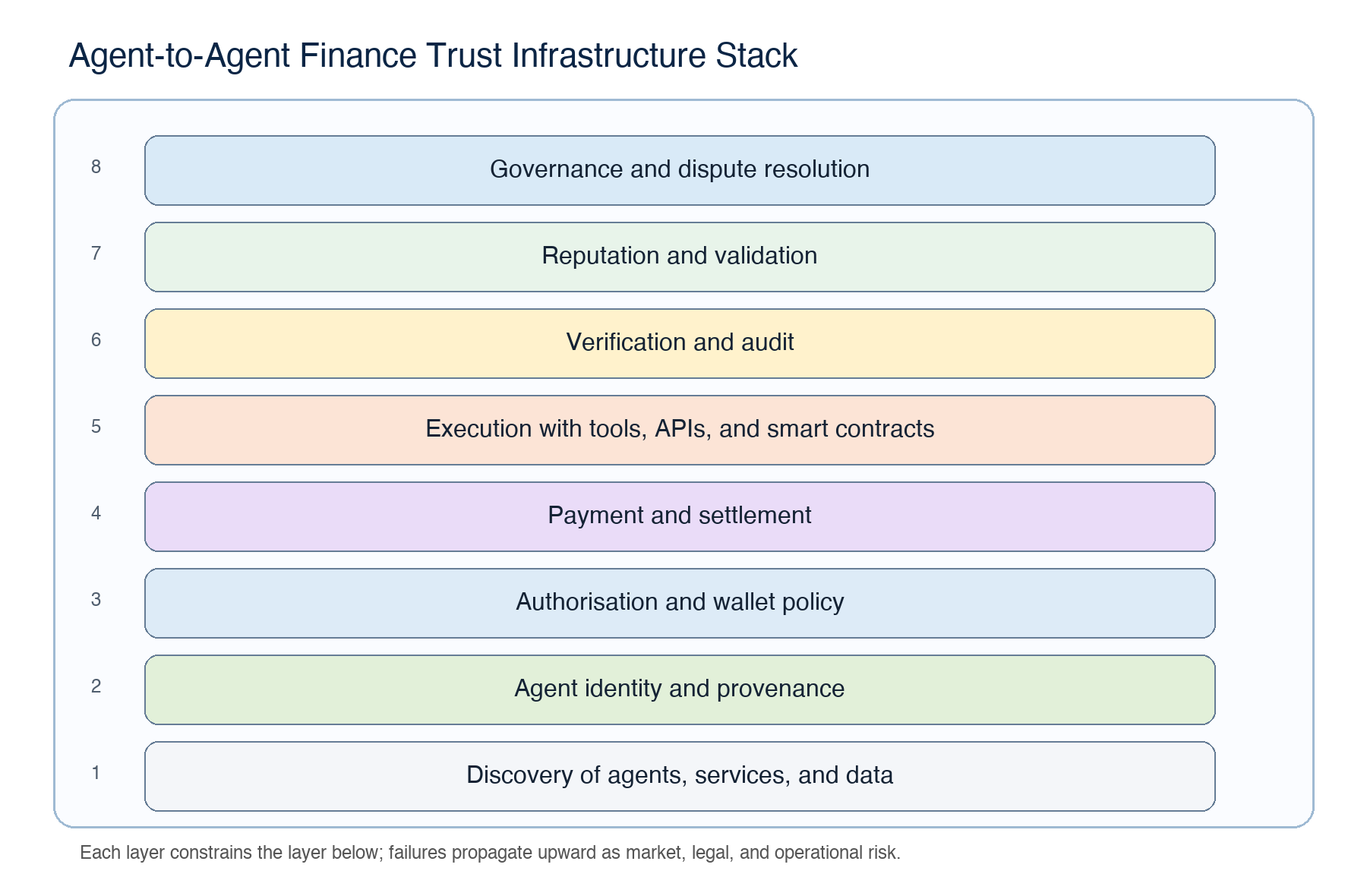}
\caption{Trust infrastructure stack for agent-to-agent finance. Source: author's elaboration.}
\label{fig:trust-stack}
\end{figure}

\subsection{Machine payments and the evidence envelope}

Agent-to-agent systems x402 and ERC-8004 are useful because they sit in different layers of the emerging agent stack. x402 belongs to the machine-payment and agent-commerce layer: it revives the HTTP 402 payment-required pattern and connects web resources to programmable stablecoin payments so that software clients, including agents, can pay for digital services. ERC-8004-style systems belong primarily to the identity, reputation and validation layer. Together they illustrate why payment and trust cannot be separated in agentic finance.

Blockchain contributes most clearly to the payment layer. x402 illustrates the emerging design pattern: a client requests a resource; the server responds with payment instructions; the client constructs a payment payload; a facilitator verifies and settles the payment; and the resource is returned if payment is valid (Coinbase Developer Platform, n.d.). This is not yet a regulated financial-market payment system, but it shows how APIs, stablecoins and machine clients can be combined into programmable service procurement. For agents, the value is not ideological decentralisation. It is the ability to pay for small digital services without opening accounts, negotiating contracts or routing every transaction through a human checkout process.

The clearest pain point in A2A payments is the mismatch between small digital services and traditional payment and procurement processes. A human institution can sign a data contract, open a vendor account and approve invoices. An agent that needs a single API call or model inference cannot wait for that process. Conversely, letting the agent use an unconstrained corporate card or private key is unacceptable. A2A payment infrastructure must therefore support low-friction transactions without abandoning control. It must make very small purchases easy while making unauthorised or suspicious purchases difficult.

Consider a concrete payment case. A risk-monitoring agent observes that a tokenised collateral position has become volatile. It requests a high-frequency price feed from an approved data agent, receives a machine-readable quote, checks that the provider is on an allowlist, confirms that the fee is below its spend cap, pays through a stablecoin rail, receives signed delivery metadata and attaches the receipt to a risk report. The payment is not the interesting part by itself. The interesting part is the chain of evidence connecting the risk event, the data request, the authority to pay, the provider identity, the payment and the resulting risk decision.

This case also shows why agent payments are different from ordinary micropayments. A human micropayment often requires only consent and delivery. An institutional agent payment requires mandate evidence, budget evidence, counterparty evidence, policy evidence and reconciliation evidence. It may also require evidence that the payment did not influence an impermissible transaction. Agentic payments therefore need richer metadata than consumer payments. The metadata may not all be public, but it must exist and be accessible to the right parties.

No common standard evidence schema has yet emerged. What is visible is a set of primitives: payment receipts, signed service-delivery metadata, wallet-policy records, agent registry entries, validation claims and audit logs. Regulated A2A finance will likely need an agentic payment evidence envelope: a structured object that binds the mandate, budget constraint, counterparty identity, policy version, quote, payment transaction, service-delivery receipt, reconciliation record and escalation outcome. The envelope need not be fully public, but it must be reconstructable by the firm and accessible to auditors or supervisors where required.

Figure 2 traces this payment and accountability flow. The process starts before the transfer of value and continues after settlement, which is why the relevant audit object is broader than the transaction hash.

\begin{figure}[htbp]
\centering
\includegraphics[width=0.96\textwidth]{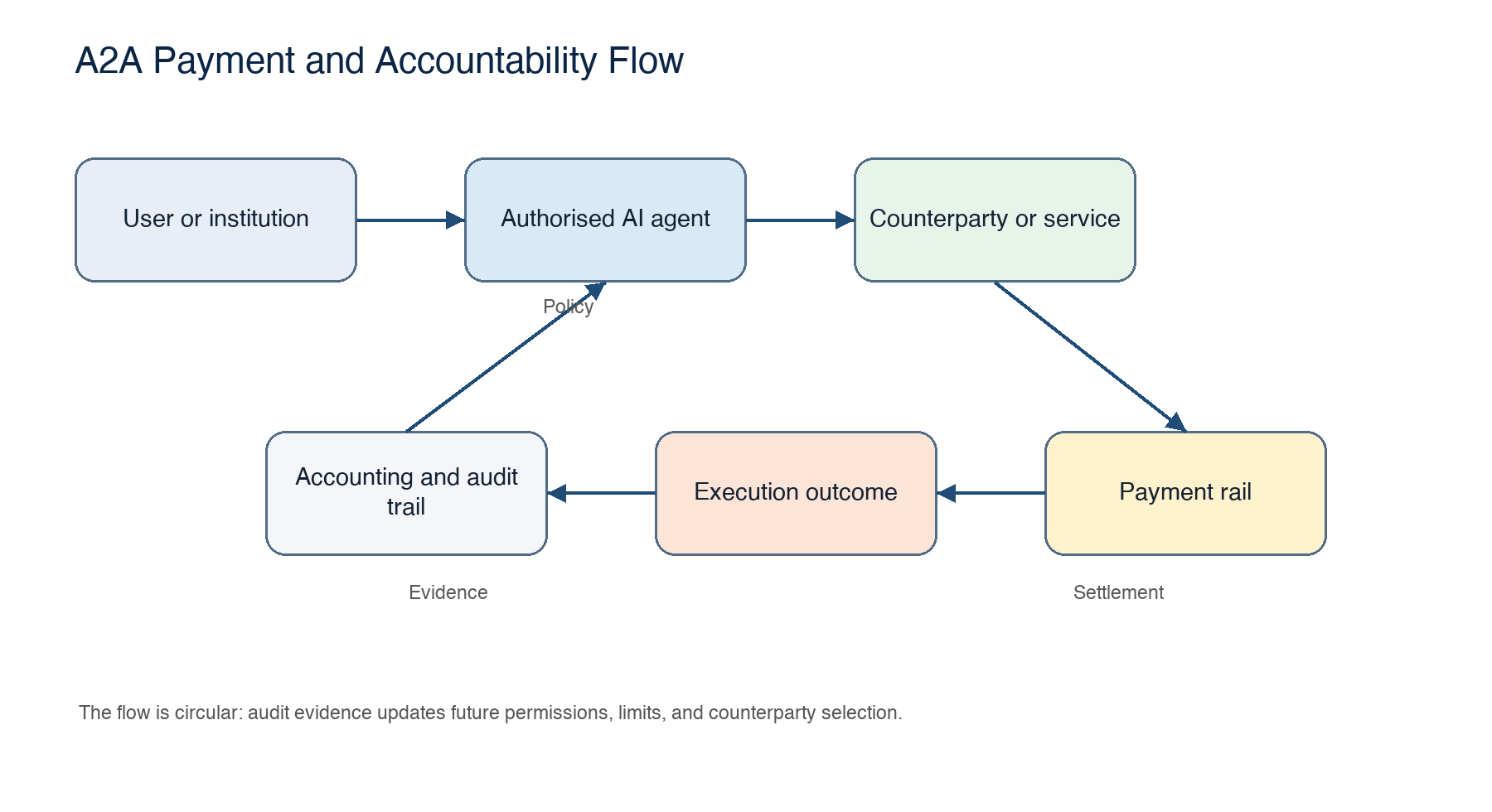}
\caption{A2A payment and accountability flow. Source: author's elaboration.}
\label{fig:payment-flow}
\end{figure}

\subsection{Wallet authorisation and policy split}

Wallets are therefore central. A standard private-key model, in which possession of the key equals unilateral control, is a poor fit for autonomous agents. Agent wallets need delegated authority rather than absolute control. They should support spending caps, time limits, asset restrictions, counterparty allowlists, emergency revocation, human approval thresholds and audit labels. Account abstraction, session keys and smart-account policy engines are relevant because they allow a principal to specify what an agent may do without handing over unconstrained power. Provenance-based subaccount systems such as PASS push this logic further by linking external actions to valid asset lineage while preserving some internal privacy (Yu et al., 2026).

The most credible regulated architecture is hybrid. Hard constraints should live as close to the wallet as possible: spending caps, asset restrictions, approved counterparties, time windows, revocation rights and session-key constraints. These controls should remain enforceable even if the orchestration layer behaves badly. Richer semantic policy, however, will often remain off-chain because it involves interpreting a mandate, judging whether a counterparty is appropriate rather than merely allowed, applying compliance context and deciding when to escalate.

This split creates the central attestation trade-off. On-chain controls are more visible and enforceable, but can be rigid, expensive and too public for institutional policies. Off-chain rules engines can handle deterministic policy questions - for example, whether a counterparty is on an allowlist, whether an amount is under a cap or whether an instrument is permitted - while preserving versioning and audit records. Semantic judgements require a different control style: confidence bands, anomaly thresholds, signed logs, policy hashes and human escalation. Zero-knowledge proofs may eventually allow a system to prove that an action satisfied a policy without exposing the policy itself, but this remains a specialised technique rather than a substitute for institutional governance.

\subsection{Identity, reputation and discovery}

ERC-8004 belongs to the identity, reputation and validation layer of the Ethereum-oriented agent ecosystem. Its promise is not payment itself, but the ability to register agents, expose service metadata, collect feedback and support validation claims. In financial terms, x402 helps an agent pay, while ERC-8004-style registries help a counterparty decide whether the paying or service-providing agent should be recognised, trusted or challenged.

Agent identity and reputation form the next layer. An agent identity should not merely be a name or wallet address. It should bind together an operator, service endpoint, model or software version, wallet authority, permitted tasks and evidence of past behaviour. ERC-8004-style registries attempt to provide identity, reputation and validation layers for decentralised AI agents. The empirical evidence is sobering. Xiong et al. (2026) find that many registrations are placeholders rather than active agents, and that trustless registries do not automatically produce trustworthy agents. Mafrur and Khusumanegara (2026) similarly find that early ERC-8004 adoption is registration-heavy but operationally shallow, with richer evidence concentrated in a small subset of agents.

The purpose of blockchain trust rails is therefore not to make trust unnecessary. It is to make specific claims auditable. A payment claim can be linked to a transaction hash. A service claim can be linked to an endpoint and signed delivery receipt. A wallet claim can be linked to a permission policy. A validation claim can be linked to a validator identity and stake or credential. These links do not settle all disputes, but they create a record that can be inspected by firms, counterparties, auditors and, where appropriate, supervisors.

\section{Verifiable, Accountable and Regulated Agentic Finance}

\subsection{Why verification changes in agentic finance}

Verification becomes necessary because agentic systems change what must be checked. In a conventional digital payment, it may be enough to verify the account, amount and recipient. In an agent-mediated payment, the economically relevant object is broader: the instruction, the interpretation of that instruction, the tool calls that led to the payment, the authority under which the agent acted, and the evidence that the purchased service affected or did not affect a downstream financial decision. Agentic finance therefore turns verification from a narrow settlement question into an end-to-end accountability question.

Verifiability is the bridge between technical execution and financial accountability. In ordinary software systems, a log may be enough to show that a function was called. In agentic finance, the relevant questions are richer: what did the agent understand the instruction to mean, which tools did it call, what information did it rely on, which policy constrained the action, why was a counterparty selected, what payment was made, and how was the outcome evaluated? The answer need not expose every model weight or chain-of-thought trace. It must provide enough evidence to reconstruct economically significant actions.

Verifiable AI is developing along several technical lines: deterministic inference, trusted execution environments, zero-knowledge proofs, signed model outputs, cryptographic commitments and challenge mechanisms. EigenAI, for example, proposes deterministic inference and a re-execution protocol so that AI outputs can be reproduced and challenged within a cryptoeconomic framework (Alves et al., 2026). For agent-to-agent finance, the broader lesson is that model outputs can be treated as claims that require evidence. A trading bot that claims to follow a certified strategy, a risk agent that produces collateral scores, or a compliance agent that validates transaction intent may need more than a natural-language explanation. It may need attestable execution conditions.

Accountability cannot be reduced to explainability. Explainability asks why a model produced an output. Accountability asks who is responsible for an action and what remedy exists when harm occurs. An agent may be explainable but unauthorised; authorised but manipulated; manipulated but still technically compliant with a narrow rule; compliant but harmful because the rule was poorly designed. Financial governance must therefore connect model governance, wallet governance, data governance and legal governance. The audit trail must show both the technical path and the mandate path.

\subsection{Regulation and supervisory observability}

The regulatory environment is moving in this direction. The EU AI Act establishes a risk-based framework for AI, with high-risk systems subject to requirements around risk management, data quality, logging, documentation, information to deployers, human oversight, robustness, cybersecurity and accuracy. The Commission states that the Act entered into force on 1 August 2024, that general-purpose AI (GPAI) obligations became applicable in August 2025, and that transparency rules come into effect in August 2026, with high-risk rules following the revised transition timetable (European Commission, 2026a). These requirements are highly relevant to agentic finance even where a particular agent is not directly classified as high-risk, because they define the emerging regulatory grammar of traceability and oversight.

Crypto and operational-resilience rules also matter. The Markets in Crypto-Assets Regulation (MiCA) creates a harmonised EU framework for crypto-assets and crypto-asset service providers, including organisational, operational, prudential, market abuse and AML expectations (European Commission, 2026b). The Digital Operational Resilience Act (DORA) focuses on digital operational resilience, information and communications technology (ICT) risk and third-party dependence in financial services. Agent-to-agent finance sits across these regimes: it may involve AI decision-making, cryptoasset settlement, digital operational dependence and outsourced services. A payment agent that uses stablecoins, third-party models and cloud-hosted tools cannot be governed by one regulatory lens alone.

The FSB's work reinforces this cross-cutting view. Its 2024 report identifies AI-related financial-sector vulnerabilities involving third-party dependencies and service provider concentration, market correlations, cyber risks, model risk, data quality and governance (FSB, 2024). Its 2026 consultation on responsible AI adoption proposes sound practices for organisation-wide AI governance and lifecycle management, explicitly asking whether practices can address emerging forms such as GenAI and agentic AI (FSB, 2026). This is a crucial point for A2A finance. Supervisors are not waiting for fully autonomous trading agents to dominate markets. They are already concerned with how complex AI systems are adopted, monitored and governed.

The hardest supervisory problem is not observing a final transaction. It is observing the relation between instruction and action. A supervisor or auditor may need to know whether a payment was connected to a legitimate mandate, whether the agent selected a counterparty through approved criteria, whether the wallet policy was changed before execution, whether a human escalation was bypassed, and whether the service output influenced a client-facing decision. These questions require structured logs, not simply blockchain explorers or model cards. They require audit objects that connect natural-language intent, technical execution and legal accountability.

A2A finance therefore sits at the intersection of AI governance, financial conduct, payments and crypto-asset regulation, operational resilience and data protection. This intersection is the reason a narrow model-risk framework is insufficient. A supervisor may need to inspect the model, the wallet policy, the counterparty-selection rule, the payment record, the outsourcing chain and the incident response process in a single workflow.

\subsection{Know Your Agent}

A practical governance model should therefore begin with Know Your Agent. KYA is not a slogan for software KYC. It is a control framework for delegated economic action. A firm should know who owns or operates an agent, which model or software version it uses, which wallet or account permissions it holds, which services it may contact, what value it may transfer, what logs it produces, which escalation rules apply, and how authority can be revoked. KYA also requires ongoing monitoring because agents may update, change tools, shift model providers or encounter new counterparties.

Know Your Agent can be decomposed into five checks. First, identity: what is the agent's persistent identifier, who operates it, and what legal person ultimately stands behind it? Second, capability: what tasks does the agent claim to perform, which model or tool stack does it use, and under what operating limits? Third, authority: which wallet, account, API keys or execution permissions does it hold, and what spending or transaction limits apply? Fourth, provenance: what evidence links an action to a user instruction, institutional policy, model version, data source and tool call? Fifth, recourse: who can pause, revoke, dispute or remediate the agent's action when something goes wrong?

The key distinction is that capability is not authority. An agent may be technically capable of routing trades, querying sanctions databases or signing transactions, but that does not mean it is permitted to do so in a particular mandate, product, jurisdiction or risk tier. Table 5 formalises Know Your Agent as a five-dimensional governance framework.

\begin{table}[htbp]
\centering
\caption{The five dimensions of Know Your Agent.}
\footnotesize
\renewcommand{\arraystretch}{1.16}
\begin{tabularx}{\textwidth}{YYY}
\toprule
KYA dimension & Governance question & Required evidence \\
\midrule
Identity & Who operates the agent and who ultimately stands behind it? & Persistent identifier, operator record, legal principal and service endpoint. \\
Capability & What can the agent technically do? & Model or software version, tool stack, declared services, testing record and operating limits. \\
Authority & What is the agent permitted to do? & Mandate, wallet scope, transaction limits, approved counterparties, policy version and revocation path. \\
Provenance & How was this action generated? & Instruction record, data source, tool call, model version, payment evidence and delivery receipt. \\
Recourse & How can the action be stopped, disputed or remedied? & Escalation rule, human owner, pause control, dispute mechanism, insurance or contractual remedy. \\
\bottomrule
\end{tabularx}
\end{table}

Figure 3 situates KYA within a wider governance map. The principal retains legal accountability, while the agent's operational conduct is bounded through identity, authority, verification, monitoring and recourse.

\begin{figure}[htbp]
\centering
\includegraphics[width=0.96\textwidth]{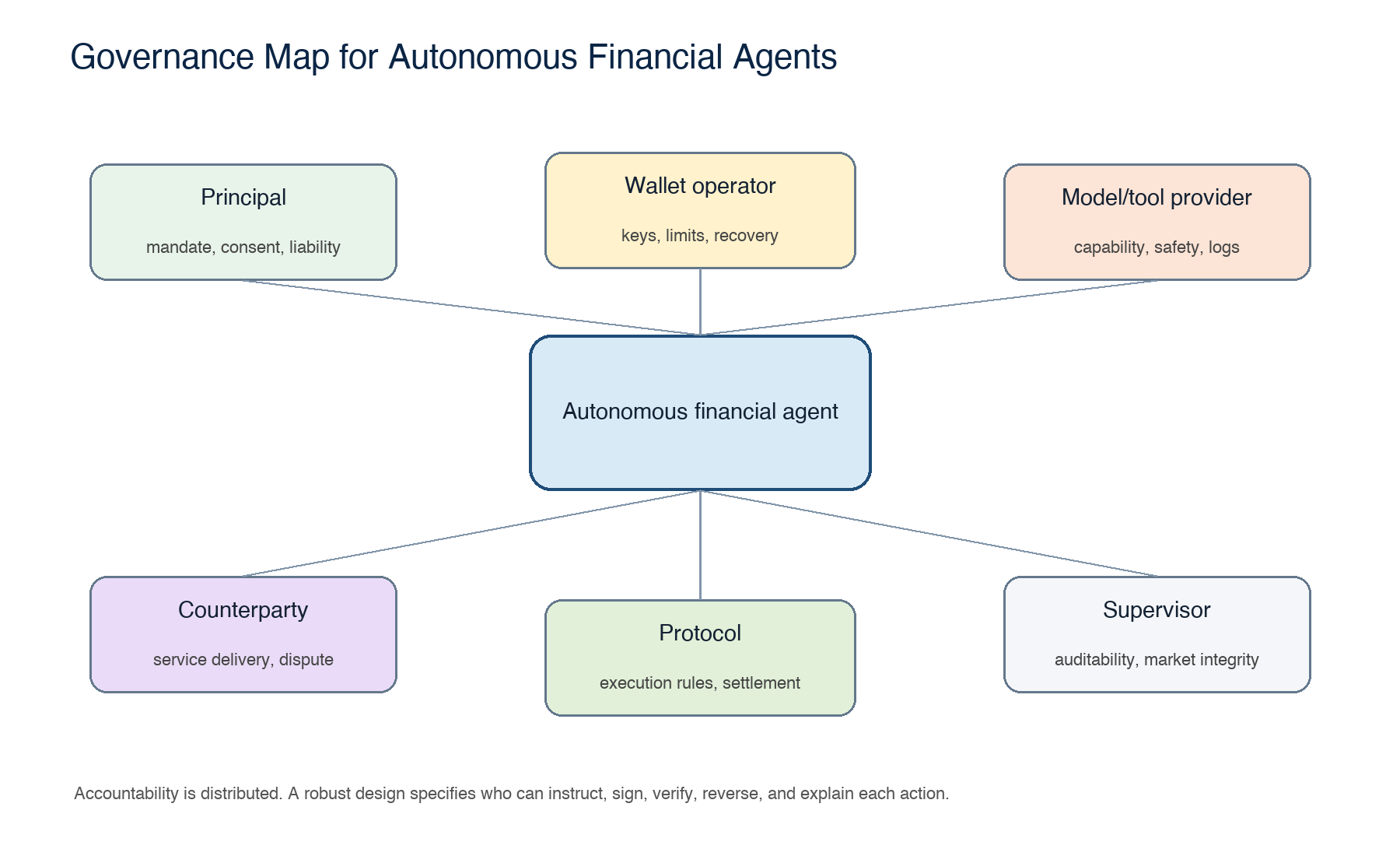}
\caption{Governance map for autonomous financial agents. Source: author's elaboration.}
\label{fig:governance-map}
\end{figure}

\subsection{Risk governance and confidence bands}

Human oversight should be risk-tiered rather than absolute. A low-value API payment for a pre-approved data query may be automated. A new counterparty, unusual payment amount, change in wallet permissions or high-impact trade should trigger escalation. A severe anomaly should revoke or pause authority. This approach is more realistic than insisting on manual approval for every action and safer than allowing unconstrained autonomy. It also aligns with financial regulation's broader logic of proportionality: controls should reflect materiality, complexity and potential harm.

Risk governance should also account for learning and updating. A deterministic script can be certified against a relatively stable code path. An adaptive agent may change behaviour as prompts, tools, memories, model versions and external services change. Model updates can alter risk even when wallet policy remains constant. Tool updates can alter execution even when the agent's mandate is unchanged. This means that governance must cover the lifecycle of agents, not only deployment approval. Testing, monitoring, version control, incident review and decommissioning become part of financial-market infrastructure.

The main risks can be grouped into seven categories. Identity risk concerns impersonation, weak registration and unclear ownership. Authorisation risk concerns agents exceeding mandates or being manipulated into unintended actions. Wallet risk concerns key management, delegated signing, session-key leakage and recovery. Payment risk concerns wrong-address transfers, failed delivery, settlement disputes and stablecoin exposure. Model and prompt risk concerns hallucinated counterparties, prompt injection, tool misuse and misinterpreted contracts. Reputation risk concerns Sybil attacks, fake feedback and collusive validation. Liability risk concerns fragmented responsibility across users, deployers, model providers, wallet operators, protocols and service counterparties.

Confidence bands are useful because the riskiest semantic judgements do not always look like obvious failures. Low-confidence actions should be escalated because the agent may not understand the mandate, counterparty or instrument. Suspiciously high-confidence actions may also deserve review if they occur in ambiguous contexts, because an agent can be manipulated into appearing certain. In regulated deployments, the confidence band should therefore be treated as a control signal rather than a model-performance ornament.

Table 6 lists seven risk channels and corresponding controls. The table deliberately separates payment and settlement risk from wallet risk because a correctly authorised wallet action can still produce a wrong-recipient transfer, failed service delivery, reconciliation gap or unstable payment exposure.

\begin{table}[htbp]
\centering
\caption{Risk channels and controls in accountable agentic finance.}
\footnotesize
\renewcommand{\arraystretch}{1.16}
\begin{tabularx}{\textwidth}{YYYY}
\toprule
Risk channel & Agentic failure & Financial consequence & Control \\
\midrule
Identity & Impersonation, placeholder registrations or unclear operator ownership. & Misattribution, counterparty fraud and weak recourse. & Persistent identities, credentials, verified endpoints and operator records. \\
Authorisation & Agent exceeds mandate or is induced to approve an unintended action. & Unauthorised payment, trade or service procurement. & Scoped permissions, thresholds, revocation, session-key limits and escalation. \\
Wallet & Session-key leakage, unconstrained signing or weak recovery. & Asset loss and operational disruption. & Smart accounts, subaccounts, recovery design and transaction simulation. \\
Payment and settlement & Wrong recipient, failed delivery, disputed service or unstable payment asset. & Irrecoverable transfer, accounting mismatch, liquidity exposure or settlement delay. & Evidence envelopes, delivery receipts, reconciliation records and stablecoin or cash controls. \\
Model and prompt & Prompt injection, hallucinated counterparty, mandate drift or manipulated confidence. & Payment-triggered harm, erroneous execution or hidden policy breach. & Tool monitoring, red-teaming, confidence bands and safe-action policies. \\
Reputation & Fake feedback, collusive validation or low-cost identities. & Mispriced trust, adverse selection and concentration. & Anti-Sybil mechanisms, multidimensional reputation and validation standards. \\
Liability & Fragmented responsibility across deployers, providers, wallets and counterparties. & Delayed remediation and supervisory uncertainty. & Contractual allocation, audit-ready evidence, incident response and recourse rights. \\
\bottomrule
\end{tabularx}
\end{table}

\section{Financial Market Applications}

Across these applications, the immediate opportunity is not to replace financial professionals. It is to automate micro-coordination. Many financial workflows involve small, repeated decisions: whether to buy a dataset, whether to call a model, whether to request a quote, whether to escalate a transaction, whether to rebalance a small position, whether to generate a compliance evidence pack. Agents can handle these decisions if their autonomy is bounded. The economic value comes from reducing latency, manual reconciliation and search costs, not from granting agents unconstrained discretion.

\subsection{Data and model procurement}

The most credible applications of agent-to-agent finance are those in which digital services are already modular, repeated and information-intensive. The first is data and model procurement. Financial firms purchase market data, alternative data, news analytics, risk scores, sanctions screening, model inference and research services. Today much of this procurement is mediated through subscriptions, enterprise contracts or static API keys. Agents make a more granular model possible: pay per query, per data slice, per inference call, per validation or per delivered artefact. x402 is an early example of a protocol designed for such programmatic payment, including AI agents paying for API access (Coinbase Developer Platform, n.d.).

The pain point in data procurement is not only payment friction. It is accountability. If a trading or risk agent buys a data signal, the firm must know why the data was purchased, whether the provider was approved, how much was paid, whether the data arrived, and whether it influenced a material decision. Agent-to-agent finance can improve this by linking payment receipts, service metadata and audit logs. The same infrastructure can support internal cost allocation: agents that consume expensive model inference or data services can be charged back to desks or strategies based on actual use.

\subsection{DeFi intent execution}

DeFi intent execution is DeFi intent execution. DeFi transactions are often opaque because user intent is hidden behind contract calls, router interactions, logs and token movements. Mao et al. (2025) propose a multi-agent LLM framework for DeFi transaction intent mining, using domain experts and evaluators to infer user intent from on-chain and off-chain data. In A2A finance, intent analysis can become operational rather than merely descriptive. An agent can formulate a swap or collateral-management intent, request quotes from solvers, evaluate execution quality, pay for routing and store an explanation of why a route was chosen.

The current feasibility of this application is strongest in controlled DeFi or institutional DeFi settings. Agents can be restricted to approved protocols, assets and transaction types. Wallet policies can cap exposure. Intent templates can limit ambiguity. Execution providers can be required to return structured evidence. The result is not a fully autonomous trader roaming across DeFi. It is a bounded execution assistant that translates economic objectives into constrained transactions and produces a record suitable for review.

\subsection{Autonomous treasury and liquidity management}

Autonomous treasury and liquidity management is autonomous treasury and liquidity management. Firms, DAOs, funds and tokenised-asset platforms increasingly manage balances across bank accounts, stablecoins, custodians, exchanges and on-chain protocols. Routine actions include paying service providers, rebalancing liquidity, monitoring collateral, sweeping idle balances, hedging stablecoin exposure and responding to margin or liquidity triggers. Agents can automate parts of this workflow if their permissions are carefully scoped. The value is operational speed and 24-hour responsiveness; the risk is that a manipulated or misconfigured agent can move real assets quickly.

\subsection{Compliance and regulatory reporting}

Compliance and regulatory reporting is compliance and regulatory reporting. Agents can collect evidence, classify transactions, screen counterparties, generate suspicious-activity explanations, monitor sanctions lists and assemble reporting packages. The link to A2A finance is that compliance itself may become a service purchased or coordinated by agents. A payment agent may call an AML screening agent before paying a new counterparty. A treasury agent may call a travel-rule or sanctions-screening service before moving stablecoins. In this setting, the compliance agent's output must be auditable, because a false negative can have legal consequences.

\subsection{Market intelligence and research}

Market intelligence and research is market intelligence and research. Research agents can pay for datasets, retrieve filings, query models, evaluate competing forecasts and produce investment memos. Here the financial transaction may be small, but the downstream decision can be material. The governance issue is provenance: which sources were purchased, which model generated which conclusion, and what evidence supports the recommendation? Agent-to-agent finance can turn research procurement into a traceable workflow rather than a set of unlogged browser or API interactions.

\subsection{Tokenised settlement and wholesale payments}

Tokenised settlement and wholesale payments concern tokenised settlement and wholesale payments. Bank for International Settlements (BIS) Project Agora demonstrates that tokenised central bank reserves and commercial bank deposits can be combined on a shared programmable platform for wholesale cross-border payments, with atomic multi-currency settlement and workflow logic embedded in transactions (Bank for International Settlements, 2026). Agent-to-agent finance can be layered on top of such infrastructure. Treasury agents could initiate conditional payments; compliance agents could check transaction prerequisites; liquidity agents could monitor settlement windows; audit agents could reconcile records. This is not the same as public-chain DeFi, but it shares the need for programmable instructions, identity, authorisation and evidence.

\subsection{Near-term implementation pathways}

One near-term implementation path is an internal agent marketplace within a regulated firm. The firm approves a set of specialised agents: a data agent, a pricing agent, a compliance agent, a treasury agent and an execution-preparation agent. Each has a service description, owner, permission scope, logging requirement and budget. Agents can request services from one another, but payments are internal accounting entries or controlled stablecoin transfers. This design allows the firm to test A2A finance without exposing clients to open public-agent markets. It also creates a template for later interaction with external agents.

A third implementation path is agent-assisted tokenised payment workflows. In wholesale or corporate payment settings, many frictions arise before settlement: checking invoice data, verifying counterparty credentials, confirming sanctions status, ensuring liquidity, applying FX rules, and reconciling post-payment records. Agents can coordinate these pre- and post-settlement tasks even when the final settlement rail is bank money or tokenised deposits rather than a public stablecoin. This matters because it shows that A2A finance is not limited to public crypto markets. The broader issue is machine-coordinated financial workflow.

Table 7 summarises these applications by executable form, infrastructure need and maturity. The point is to distinguish current deployable workflows from longer-term market-infrastructure visions.

\begin{table}[htbp]
\centering
\caption{A2A finance application maturity map.}
\footnotesize
\renewcommand{\arraystretch}{1.16}
\begin{tabularx}{\textwidth}{YYYY}
\toprule
Application & Current executable form & Infrastructure need & Maturity \\
\midrule
Data and model procurement & Agents pay for application programming interface calls, datasets and inference on a usage basis. & Programmatic payments, receipts, provenance and cost allocation. & Near-term. \\
Internal agent marketplace & Approved agents request services from other approved agents inside a firm. & Service registry, budgets, internal ledger entries, human ownership and audit logs. & Near-term. \\
Compliance services & Agents call screening, reporting or evidence-assembly services before payment or settlement. & Signed outputs, records, escalation rules and legal review. & Near-term to emerging. \\
DeFi intent execution & Agents formulate constrained intents and request solver quotes. & Wallet policy, intent binding, execution evidence and maximal extractable value (MEV) controls. & Emerging. \\
Autonomous treasury & Agents rebalance liquidity and make routine payments under caps. & Smart accounts, spend limits, counterparty lists and anomaly monitoring. & Emerging. \\
Prediction and oracle systems & Agents submit evidence, challenge claims or pay for verification. & Validation standards, dispute mechanisms and anti-collusion design. & Emerging. \\
Tokenised wholesale settlement & Agents coordinate conditional payments and reconciliation on programmable platforms. & Identity, legal finality, compliance triggers and regulated settlement rails. & Longer-term, infrastructure-dependent. \\
\bottomrule
\end{tabularx}
\end{table}

\section{Conclusion: Designing Bounded Autonomy}

Agent-to-agent finance names a shift in the operational structure of financial markets. AI agents are not merely improving forecasts or summarising documents; they are beginning to coordinate services, make payments, express intent and interact with transaction infrastructure. This does not mean that autonomous agents will immediately become independent market participants in a legal sense. It means that delegated software action will become more economically consequential, and market infrastructure must be designed accordingly.

The chapter has argued that the central design problem is bounded autonomy. The relevant question is not whether agents can transact, but under what constraints they should transact. A trustworthy A2A financial system must identify agents, bind them to principals, limit wallet authority, verify service delivery, record payment intent, monitor reputation and preserve audit evidence. Blockchain can help with some of these functions by providing programmable settlement, persistent identifiers and auditable state. It cannot replace institutional governance, legal responsibility or supervisory judgement.

The conclusion is therefore not that every financial agent should be on-chain, nor that every transaction should be automated. It is that financial markets need a language and infrastructure for delegated software action. The move from AI-assisted finance to agentic finance changes the object of governance from model output to agent conduct. A model may predict; an agent may act. That difference is the reason identity, payment, verification and accountability must be designed together.

The chapter's research agenda can be organised into four programmes. The first concerns identity and delegated authority: how agent identities should bind operators, models, wallets, service endpoints and legal principals; how Know Your Agent can become operational; and how capability should be separated from authority across changing mandates.

The second programme concerns market infrastructure and machine payments. Researchers need to examine which rails can support low-value, high-frequency machine payments while preserving compliance, dispute resolution, resilience and reconciliation. This includes x402-style service payments, internal agent ledgers, stablecoin settlement, tokenised deposits and future wholesale settlement platforms.

The third programme concerns verification, reputation and accountability. Agentic finance needs methods for proving service delivery, evaluating reputation, resisting Sybil attacks, auditing payment intent and allocating liability across principals, deployers, model providers, wallet operators and counterparties. The challenge is selective verifiability: enough evidence for accountability without exposing every sensitive policy or client detail.

The best way forward is neither promotional enthusiasm nor defensive rejection. Autonomous financial agents are likely to be adopted where they solve concrete workflow problems: data procurement, model access, DeFi intent execution, treasury operations, compliance evidence and tokenised payment workflows. The task for researchers, institutions and regulators is to design the trust infrastructure before the most fragile forms of autonomy become deeply embedded. Agent-to-agent finance should therefore be understood as an invitation to rethink financial market infrastructure for a world in which software agents can act, but must remain accountable.

\section*{Acknowledgements}

I thank Liam Johnston for helpful comments on authorisation and evidence design.

\section*{References}
\begin{list}{}{%
\setlength{\leftmargin}{0.28in}%
\setlength{\itemindent}{-0.28in}%
\setlength{\itemsep}{0.35em}%
\setlength{\parsep}{0pt}}
\item Alqithami, S. (2026) 'Autonomous Agents on Blockchains: Standards, Execution Models, and Trust Boundaries', arXiv:2601.04583.
\item Alves, D.R., Patankar, V., Pereira, M., Stephens, J., Vaziri, N. and Kannan, S. (2026) 'EigenAI: Deterministic Inference, Verifiable Results', arXiv:2602.00182.
\item Arner, D.W., Barberis, J. and Buckley, R.P. (2017) 'FinTech, RegTech, and the reconceptualization of financial regulation', Northwestern Journal of International Law and Business, 37(3), pp. 371-413.
\item Bank for International Settlements (2026) 'Project Agora: exploring tokenisation of wholesale cross-border payments'. Available at: \url{https://www.bis.org/about/bisih/topics/fmis/agora.htm} (Accessed: 30 June 2026).
\item Bank of England and Financial Conduct Authority (2024) 'Artificial intelligence in UK financial services - 2024'. Available at: \url{https://www.bankofengland.co.uk/report/2024/artificial-intelligence-in-uk-financial-services-2024} (Accessed: 30 June 2026).
\item Coase, R.H. (1937) 'The nature of the firm', Economica, 4(16), pp. 386-405.
\item Coinbase Developer Platform (n.d.) 'x402: Overview'. Available at: \url{https://docs.cdp.coinbase.com/x402/} (Accessed: 30 June 2026).
\item Committee on Payments and Market Infrastructures and International Organization of Securities Commissions (2012) Principles for financial market infrastructures. Basel: Bank for International Settlements.
\item Eisenhardt, K.M. (1989) 'Agency theory: an assessment and review', Academy of Management Review, 14(1), pp. 57-74.
\item European Commission (2026a) 'AI Act'. Available at: \url{https://digital-strategy.ec.europa.eu/en/policies/regulatory-framework-ai} (Accessed: 30 June 2026).
\item European Commission (2026b) 'Crypto-assets'. Available at: \url{https://finance.ec.europa.eu/digital-finance/crypto-assets_en} (Accessed: 30 June 2026).
\item European Parliament and Council (2022) Regulation (EU) 2022/2554 on digital operational resilience for the financial sector.
\item European Parliament and Council (2023) Regulation (EU) 2023/1114 on markets in crypto-assets.
\item European Parliament and Council (2024) Regulation (EU) 2024/1689 laying down harmonised rules on artificial intelligence.
\item Financial Stability Board (2024) 'The Financial Stability Implications of Artificial Intelligence'. Available at: \url{https://www.fsb.org/2024/11/the-financial-stability-implications-of-artificial-intelligence/} (Accessed: 30 June 2026).
\item Financial Stability Board (2026) 'Sound Practices for Responsible Adoption of Artificial Intelligence (AI): Consultation report'. Available at: \url{https://www.fsb.org/2026/06/sound-practices-for-responsible-adoption-of-artificial-intelligence-ai-consultation-report/} (Accessed: 30 June 2026).
\item Google Developers Blog (2025) 'Announcing the Agent2Agent Protocol (A2A)'. Available at: \url{https://developers.googleblog.com/en/a2a-a-new-era-of-agent-interoperability/} (Accessed: 30 June 2026).
\item Jensen, M.C. and Meckling, W.H. (1976) 'Theory of the firm: managerial behavior, agency costs and ownership structure', Journal of Financial Economics, 3(4), pp. 305-360.
\item Jin, Y., Wu, S., Chen, C., Bao, L., Yang, X. and Chen, J. (2026) 'The Web4 Agent Economy: A Large-Scale Empirical Study of the Landscape, Challenges, and Opportunities', arXiv:2606.25876.
\item Mafrur, R. and Khusumanegara, P. (2026) 'From Agent Identity to Agent Economy: Measuring the Operational Readiness of ERC-8004 AI Agents', arXiv:2606.12128.
\item Mao, Q., Zhang, Y., Chen, J., Zhou, W. and Yan, J. (2025) 'Know Your Intent: An Autonomous Multi-Perspective LLM Agent Framework for DeFi User Transaction Intent Mining', arXiv:2511.15456.
\item Merton, R.C. and Bodie, Z. (1995) 'A conceptual framework for analyzing the financial environment', in Crane, D.B. et al. (eds) The Global Financial System: A Functional Perspective. Boston: Harvard Business School Press, pp. 3-31.
\item Nannini, L. et al. (2026) 'AI Agents under EU Law: A Compliance Architecture for AI Providers', arXiv:2604.04604.
\item Williamson, O.E. (1979) 'Transaction-cost economics: the governance of contractual relations', Journal of Law and Economics, 22(2), pp. 233-261.
\item Xiong, X., Li, Z., Wei, W., Wang, Q., Knottenbelt, W. and Wang, Z. (2026) 'Can Trustless Agents Be Trusted? An Empirical Study of the ERC-8004 Decentralized AI Agent Ecosystem', arXiv:2606.26028.
\item Yu, J., Zhou, S., Yin, H. and Seong, B. (2026) 'PASS: A Provenanced Access Subaccount System for Blockchain Wallets', arXiv:2604.22602.
\item Zeng, Q. et al. (2026) 'When AI Meets Wall Street: A Survey on Trustworthy AI in Fintech', arXiv:2605.30650.
\item Zetzsche, D.A., Arner, D.W. and Buckley, R.P. (2020) 'Decentralized finance', Journal of Financial Regulation, 6(2), pp. 172-203.
\item Zhang, Y., Xiang, Y., Lei, Y., Wang, Q., Qiu, T., Sun, Y., Zarkov, S., Yuen, T.H., Deppeler, A., Yu, J. and Lam, K.-Y. (2026) 'SoK: Blockchain Agent-to-Agent Payments', arXiv:2604.03733.
\end{list}

\end{document}